\documentclass{article}

\usepackage[utf8]{inputenc} 
\usepackage[T1]{fontenc}    
\usepackage{hyperref}       
\usepackage{url}            
\usepackage{booktabs}       
\usepackage{amsfonts}       
\usepackage{nicefrac}       
\usepackage{microtype}      
\usepackage{algorithm}
\usepackage{algpseudocode}
\usepackage{subcaption}
\usepackage{enumitem}
\usepackage{caption}

\usepackage{bbm}
\usepackage{amsmath}
\usepackage{amssymb}
\usepackage{amsthm}

\usepackage{graphicx}
\usepackage{epstopdf} 
 
\usepackage{multirow}

\usepackage{times}
\usepackage{braket} 
\newtheorem{thm}{Theorem}

\newtheorem{defn}{Definition}

\newcommand{\algo}{CANN}

\newcommand{\bX}{\mathbf{X}}
\newcommand{\bXt}{\mathbf{X}_t}

\newcommand{\bbt}{\mathbf{b}_t}
\newcommand{\bb}{\mathbf{b}}

\newcommand{\nr}{\mathbb{R}}
	\newcommand{\nf}{\mathcal{F}}

	\newcommand{\np}{\mathbb{P}}
	
	\newcommand{\nE}{\mathbb{E}}
	\newcommand{\lamoptim}{\lambda_\infty^*}

    \newcommand{\Strategy}{\mathcal{S}}
    \newcommand{\strategy}{\mathbf{S}}

    \newcommand{\choS}{\mathcal{B}} 
    \newcommand{\obs}{X} 
    \newcommand{\obsr}{x} 
    \newcommand{\cho}{{(\mathbf{b},c)}} 
    \newcommand{\obsF}{X} 
    \newcommand{\optimal}{\mathcal{W}^*} 
    \newcommand{\stri}{w}

\newcommand{\eqdef}{\triangleq}

\title{  Growth-Optimal Portfolio Selection under CVaR Constraints}

\author{Guy Uziel  \\ 
Technion -- Israel Institute of Technology  \\
Ran El-Yaniv  \\ 
Technion -- Israel Institute of Technology  \\
}

\begin{document}

\maketitle

\begin{abstract}
 Online  portfolio selection research has so  far focused mainly on minimizing regret 
defined in terms of wealth growth. Practical financial decision making, however, is deeply concerned with 
both wealth and risk. 
We consider online learning of portfolios of stocks whose prices are governed by arbitrary (unknown) stationary and ergodic processes, where the goal is to maximize wealth while keeping the conditional value at risk (CVaR) below a desired threshold.
We characterize the asymptomatically optimal risk-adjusted performance and  present an investment strategy whose portfolios are guaranteed to achieve the asymptotic optimal solution while  fulfilling  the desired risk 
constraint. We also numerically demonstrate and validate the viability of our method on standard datasets.
\end{abstract}

\section{Introduction}
\label{sec:intro}

It has long been recognized that the value of any financial investment should 
be quantified using both return and risk, where risk is traditionally measured by the variance of the return.
A common quantification for risk-adjusted return is the Sharpe ratio 
\cite{sharpe1975adjusting}, which is essentially the (annualized) mean return divided by the (annualized)
standard deviation of the return.
Nevertheless, in online \emph{portfolio selection} \cite{Cover1991}, which has become a focal point in online learning research, risk is rarely considered and the primary 
quantity to be optimized is still the return alone. The creation of an online learning technique 
that optimizes risk-adjusted return is a longstanding goal and a major challenge
\cite{LiH2014}.

In an adversarial (regret minimization) online learning setting,
 risk-adjusted portfolio selection with no regret is known to be an impossible goal
\cite{EvenKW2006,MannorTY2009}. Recently, within an i.i.d. setting,  Mahdavi et al.  presented a framework that can be utilized for achieving this goal  \cite{MahdaviYJ2013}, and Haskell et al. considered risk-aware algorithms \cite{HaskellXCZ2016}, but i.i.d. modeling has been criticized for being unsuitable for  modeling the stock prices faithfully 
\cite{LoM2002}.
The problem with i.i.d. modeling is the lack of time dependencies between stock returns.
A substantially richer family of stochastic models is the class of stationary and ergodic 
processes, which are sufficiently expressive to model arbitrary dependencies among stock prices.

Many publications have considered stationary and ergodic markets 
 \cite{Algoet1988,GyorfiUW2008,GyorfiS2003,GyorfiLU2006,LiHG2011}, and  all
these works consider strategies that are oblivious to risk. 
Moreover, all the learning strategies they consider rely on non-parametric estimation 
techniques (e.g., histogram, kernel, or nearest neighbors methods). Moreover,
these strategies always use a countably infinite set of experts, and the 
guarantees provided for these strategies are always asymptotic. 
This is no coincidence, as  it is well known that finite sample guarantees for these methods cannot be achieved without
additional strong assumptions on the source distribution \cite{DevroyeGG2013,VonS2008}. 
Similarly, it is also known that non-parametric strategies in this context must rely on infinitely many 
experts \cite{GyorfiLM1999}.

Approximate implementations of non-parametric strategies (which apply only a finite set of experts), however, turn out to work exceptionally well and, despite   the inevitable approximation, are reported
\cite{GyorfiUW2008,GyorfiS2003,GyorfiLU2006,LiH2012,LiH2014}
to significantly outperform
 strategies designed to 
work in an adversarial, no-regret setting.
For example, the nearest-neighbor investment strategy of \cite{GyorfiUW2008}
is shown in \cite{li2015online,LiH2014} to beat
Cover's universal portfolios (UP) \cite{Cover1991}, 
the exponentiated gradient (EG) method \cite{HelmboldSSW1998}, and the online Newton steps strategy
of \cite{AgarwalHKS2006}  on most of the common datasets. 
We also note that  practical approximate use of asymptotic methods is prevalent in  other areas of machine learning such as 
(deep) reinforcement learning with function approximation \cite{bhatnagar2009convergent}).

For a market with $n$ stocks, and 
within a stochastic online learning framework, we 
develop a novel online portfolio selection strategy called \emph{CVaR-Adjusted Nearest Neighbor} 
(\algo),
which guarantees the best possible asymptotic performance while keeping the risk contained
to a desired threshold.
This is done using a novel mechanism that facilitates the handling of multiple objectives.
Rather than using standard deviation to measure risk, we consider the 
 well-known CVaR, a coherent and widely-accepted risk measure, which improves upon  the traditional measure by appropriately capturing the downside risk \cite{RockafellarU2000}.
We prove the asymptotic optimality of our strategy for general stationary and ergodic
processes, thus allowing for arbitrary (unknown) dependencies among stock prices. 
We also present numerical examples where we apply an approximate application of our strategy
(with a finite set of experts) that validates the method and beautifully demonstrates how risk can be controlled.


%

 \section{Online Portfolio Selection }
 \label{sec:OPS}
 We consider the following standard online portfolio selection game with short selling and leverage, as 
 defined  by Gy{\"o}rfi et al. \cite{GyorfiW2012}.
 The game is
 played through $T$ days over a market with $n$ stocks.
 On each day $t$, the market is represented by
 a \emph{market vector} $\bX_t$ of relative prices, 
 $\bX_t \eqdef (x_{1}^{t},x_{2}^{t},...,x_{n}^{t})$,
 where for each $i=1, \ldots ,n$, 
 $x_{i}^{t}\geq0$ is the \emph{relative price} of stock $i$, defined to be the ratio of its closing price on day $t$ relative to its closing price on day	 $t-1$. 
  A \emph{wealth allocation} vector or \emph{portfolio} for day $t$ is 
 $\bbt \eqdef (b_{0}^{t},b_{1}^{t}, b_{2}^{t} , \ldots, b_{n+1}^{t})$, where $b_{0}^{t}$ is a cash allocation (not invested in any stock), and for $i>0$, $b_{i}^{t}$ is the wealth allocation for stock $i$, where a positive 
 component, $b_{i}^{t} > 0$, represents a \emph{long position} in stock $i$, and a negative one, $b_{i}^{t} < 0$, is a \emph{short position} in stock $i$.
 We also allow leverage; that is, the investor can borrow and invest additional 
 cash, so as to amplify her profits. For  the borrowed cash, the investor must pay a daily   interest rate, $r>0$, and we assume that the investor receives the same interest $r$ for deposited cash ($b_0^t$).  
 Consider a portfolio $\bbt$ played at the start of day $t$. 
 After the market vector $\bXt$ is revealed, the portfolio changes in response to changes in stock price,  as follows.
 For each portfolio component $b_i$,
 if  $b_{i}^{t}>0$ is a long position, its revised value is $b_{i}^{t}x_{i}^{t}$. However, 
 if $b_{i}^{t}<0$ is a short position, then, after we take into account the interest owed on borrowing the stock for the short sale, the revised value of this position is  
 $b_{i}^{t}(x_{i}^{t}-1+r)$ (note that in this case, the investor profits when the price drops and vice versa). 
 Clearly, short selling and leveraging are risky; for example, a short position has unbounded potential loss that is further amplified by leveraging.
 Following \cite{GyorfiW2012}, we assume that no stock can lose or gain more than $B\times100 \%$
 of   its value from one day to another, where $B \in (0,1)$. 
 In other words, for each $i, t$, 
 \begin{equation}
 \label{eq:boundedness}
 1-B \leq  x^t_i \leq 1+B.
 \end{equation}
 The allowed leverage is thus
  $L_{B,r} \eqdef\frac{B+1}{r+1}$, which is chosen to preclude the possibility of bankruptcy 
  (see, e.g.,  \cite{GyorfiW2012}, Chapter~4). 

Using the notation 
$$(\bb)^+ \eqdef ( \max\{b_1,0\},\ldots,\max\{b_n,0\})$$ and 
$$(\bb)^- \eqdef ( \min\{b_1,0\},\ldots,\min\{b_n,0\}),$$  and considering the 
interest accredited for deposited cash, the interest debited for borrowed stocks (short positions), 
and the interest  paid for leveraged wealth,
we obtain, by the end of the day, an overall daily return of 
\begin{gather}
b_0(1+r)+\left\langle (\bb_t)^+,\bX_t \right\rangle + \left\langle (\bb_t)^-,\bX_t-1+r \right\rangle  - (L_{B,r}-1)(1+r) .
\end{gather}

The investor  chooses a portfolio from the following set,
\begin{gather}
\left\{ (b_0,\ldots,b_n)\in \nr^n \ \ \ \mid \ \ \ \sum_{i=1}^n |b_i| = L_{B,r} \right\},
\end{gather}
which is, unfortunately, not  convex. We thus apply a simple transformation proposed by Gy{\"o}rfi et al. \cite{GyorfiW2012}:  
transform  the market vector $\bXt$ into a vector with $2n+1$ entries (one entry for cash, 
$n$ entries for the long components, and $n$ for the short ones). Formally, we define the transformed
market vector as
 \begin{gather*}
 \bX'_t \eqdef (1+r,x^t_1,2-x^t_1+r,\ldots ,x^t_n,2-x^t_n+r),
 \end{gather*}
 which is uniquely defined as a function of the original market vector.
 The transformed portfolio set is now defined as 
 \begin{gather}
\mathcal{B}'  \eqdef \{ (b_0,\ldots,b_{2m})\in \nr^{2n+1} \ \ \mid \ \ b_i\geq 0 , \sum_{i=1}^n b_i = L_{B,r} \},
 \end{gather}
which is an unnormalized simplex. 
With this transformed market vector and portfolio set, at the  start of each trading day $t$, the player
 chooses a portfolio $\bb_t \in \mathcal{B}'$ based on  the previous market sequences. It 
 can easily be shown \cite{GyorfiW2012} that by the end of day $t$, the player's daily multiplicative return is
 simplified to
\begin{gather}
\label{eq:loss}
\left\langle \bbt ,\bX'_t \right\rangle-(L_{B,r}-1)(1+r).
\end{gather} 
With respect to a fixed stationary and ergodic process, we
denote by $\bX \eqdef \{\bXt\}_{-\infty}^\infty$\footnote{By Kolmogorov's extension theorem
	\cite{Breiman1992}, the stationary and ergodic process $(X_n)^\infty_{1}$ can be extended to $(X_n)^\infty_{-\infty}$ such that the ergodicity holds for both $n\rightarrow \infty$ and $n\rightarrow -\infty$.} 
the induced sequence of stationary and ergodic market vectors, and define the player's \emph{investment strategy}, denoted by $\strategy$, as a sequence of portfolios $  \bb_1,\bb_2, \ldots$.
 Then, assuming initial wealth of \$1, we obtain after   $T$ days the following cumulative wealth,
 \begin{equation}
 R_T (\strategy,\bX)  \eqdef  \prod_{t=1}^{T}
 \left(\left\langle \bbt ,\bX'_t \right\rangle-(L_{B,r}-1)(1+r)\right).
 \end{equation}

Defining the \emph{average growth rate},
\begin{equation}
\label{eq:W}
W_T(\strategy) \eqdef \frac{1}{T}\sum_{t=1}^{T}\log \left(\left\langle \bbt ,\bX'_t \right\rangle-(L_{B,r}-1)(1+r)\right),
\end{equation}
we have
\begin{equation*}
R_T (\strategy, \bX)  =  \prod_{t=1}^{T}\left\langle \bbt ,\bXt \right\rangle = e^{\sum_{t=1}^{T}\log \left(\left\langle \bbt ,\bXt \right\rangle-(L_{B,r}-1)(1+r)\right)}=e^{TW_T(\strategy)} .
\end{equation*}
Notice that maximizing $W_T(\strategy)$ is equivalent to maximizing $R_T (\strategy, \bX)$. 
In Section~\ref{sec:optimal} , we denote the summand of $W_T(\strategy)$ (\ref{eq:W}) by
\begin{equation}
\omega(\bbt ,\bXt ) \eqdef -\log \left(\left\langle \bbt ,\hat{\bX}_t \right\rangle-(L_{B,r}-1)(1+r)\right).
\end{equation}


\newcommand{\CVaR}{\text{CVaR}}
\newcommand{\cvar}{\text{CVaR}}

 \section{Introducing Risk} 

The traditional quantity for measuring financial risk is the variance (standard deviation) of the return. This measure, however, is criticized for  being inadequate to measure risk. One of the reasons is its inability to distinguish between downside risk and upside risk (which corresponds to a desirable behavior). Various alternative measures have been proposed, such as the maximum drawdown,
and value at risk (VaR).
An axiomatic approach proposed by Artzner et al.  \cite{ArtznerDEH1999} identifies
\emph{coherent} risk measures, which satisfy the proposed axioms.
 Accordingly, the most popular coherent risk measure is \emph{conditional value at risk} (CVaR).
For any parameter $\alpha \in (0,1)$, $\cvar_\alpha$ is essentially the average loss that the investor suffers on the $(1-\alpha)\%$ worst returns. For a continuous, bounded mean random variable $Z$ the $\cvar_\alpha$ is defined as

%
%


\begin{defn}[CVaR$_\alpha$]
Let $Z$ be a continuous random variable representing loss. Given a parameter $0 < \alpha < 1$, the
CVaR$_\alpha$ of $Z$ is
\begin{gather*}  
\CVaR_{\alpha} (Z) = \mathbb{E}[Z \mid Z \geq \min\{c \mid \mathbb{P}_Z(Z\leq c) \geq \alpha \}]. 
\end{gather*}
\end{defn}

Assuming that we already know the distribution of returns, a direct calculation of CVaR from the above formula requires a calculation of the  $(1-\alpha)\%$ quantile followed by averaging over the left tail. 
Alternatively, it was shown in \cite{RockafellarU2000}  that $\CVaR_\alpha$ can be 
computed by solving the following convex optimization problem. Define
\begin{gather}
\phi'(\bb,c) \eqdef c+\frac{1}{1-\alpha}\nE\left[\left(-\log(\left\langle \bb ,\bX \right\rangle)-c\right)^+\right], \label{eq:RUfor}
\end{gather}
where we overload the previously defined $(\cdot)^+$ for vectors, and define for any scalar $x$,  
$(x)^+ \eqdef \max\{0,x\}$.

\begin{thm}[\cite{RockafellarU2000}]
\label{thm:Rock}
The function $\phi'(\bb,c)$ is convex and continuously differentiable. Moreover, the CVaR$_{\alpha}$ of the loss associated with any portfolio $\bb$ is
\begin{gather}
 \CVaR_\alpha  (\bb)= \min_{c\in\nr} \phi'(\bb,c). \label{eq:cvar}
\end{gather}
\end{thm}
Theorem~\ref{thm:Rock} is essential to the development and analysis of our strategy. 
By our market boundedness assumption (\ref{eq:boundedness}),  it follows that $\omega(\bb,\obs)$ is  contained in $[-M,M]$ for some $M>0$. Thus, any $c$ that minimizes Equation~(\ref{eq:cvar})
must reside in $[-M,M]$. 
For a complete proof of this simple fact, see \cite{HaskellXCZ2016}. 
In Section \ref{sec:optimal}, we require the following definition,
\begin{gather*}
\mathcal{B} \eqdef \mathcal{B}' \times [-M,M].
\end{gather*}

\section{Optimality of $\optimal$}
\label{sec:optimal} 
   Let $\nf_{\infty}$ be the $\sigma$-algebra generated by the infinite past $\obs_{-1},\obs_{-2},\ldots$,
   and let  $\np_\infty$, be the induced 
   regular conditional probability distribution of $\obsF_0$  given the infinite past.
   Thus, all expectations w.r.t. $\obsF_0$ are conditional given the infinite past.
    A well-known result appearing in \cite{Algoet1988,Algoet1994}  proves the following upper bound on the asymptotic average growth rate of any investment strategy $\strategy$ under stationary and ergodic markets: 
 \begin{gather}
 \label{eq:optimalGrowthRate}
\limsup_{T \rightarrow \infty} W_T(\strategy) \leq \nE \left[ \max_{\bb \in \mathcal{B}'} \nE_{\np_\infty} \left[-\omega( \bb ,\bX_0) \right] \right] .
 \end{gather}  
 Over the years, several algorithms achieving this asymptotic bound were proposed  \cite{GyorfiS2003,GyorfiLU2006,GyorfiUW2008} (for the case of long-only portfolios).

 Our goal is to achieve the optimal asymptotic average growth rate
 while keeping the CVaR bounded. 
 By Theorem~\ref{thm:Rock},  the desired  growth rate is given by the solution to the following minimization problem,
 \begin{equation}
\label{minprob}
\begin{aligned}
& \underset{(\bb,c) \in \choS}{\text{minimize}}
& &  \nE_{\np_\infty} [ \omega( \bb ,\bX_0) ] \\
& \text{subject to}
& & \phi(\bb,c) \leq \gamma, 
\end{aligned}
\end{equation} 
where 
\begin{gather*}
\phi(\bb,c) \eqdef c+\frac{1}{1-\alpha}\nE_{\np_\infty}\left[\left(-\log(\left\langle \bb ,\bX \right\rangle)-c\right)^+\right].
\end{gather*}
Optimization problem (\ref{minprob}) motivates a definition of  a \emph{$\gamma$-bounded strategy}, 
whose long-term  average CVaR, calculated according to  the available information at the beginning of each round, is bounded by $\gamma$.
\begin{defn} [$\gamma$-bounded strategy]
An investment strategy $\strategy$ will be called $\gamma$-bounded if, almost surely,
\begin{gather*}
\limsup_{T\rightarrow\infty} \frac{1}{T}\sum_{i=1}^{T} \min_{c \in \nr}\left(  c+\frac{1}{1-\alpha}
\nE_{\np_{X_i\mid X_0^{i-1}}}\left[\left(-\log(\left\langle \bb ,\bX \right\rangle)-c\right)^+\right] \right) \leq \gamma.  
\end{gather*}
 The set of all  $\gamma$-bounded strategies   is denoted $\Strategy_\gamma$.
\end{defn} 
Clearly, there is always a solution to optimization problem  (\ref{minprob}), and therefore,
$\Strategy_\gamma \neq \emptyset$. For example, the vacuous strategy that always invests everything
in cash  is $\gamma$-bounded for any $\gamma>0$.
Let $(\bb_{\infty}^*,c_{\infty}^* )$ be a solution to  (\ref{minprob}).
Define the \emph{$\gamma$-feasible optimal value} as
\begin{gather*}
\optimal  \eqdef \nE\left[ \nE_{\np_\infty}\left[\omega (\bb_{\infty}^*,\obs_0)\right]\right] \phantom{a} a.s. 
\end{gather*}

Optimization problem~(\ref{minprob}) is convex over $\choS$, which 
in turn is a compact and convex subset of $\nr^{2n+2}$. Therefore, the problem is equivalent to finding  the saddle-point of the Lagrangian function \cite{BenN2012}, namely,
\begin{gather}
\label{eq:minMax}
\min_{(\bb,c) \in \choS}\max_{\lambda \in \nr^+}\mathcal{L}(\cho,\lambda),
\end{gather}
where the Lagrangian is
\begin{gather}
\label{eq:Lagrangian}
\mathcal{L}(\cho,\lambda)\eqdef  \nE_{\np_\infty}\left[\omega (\bb,\obs_0)\right]+\lambda\left( \phi (\bb,c)-\gamma\right).
\end{gather}
Let $\lamoptim$ be the value of $\gamma$ optimizing (\ref{eq:minMax}), and 
assume it is unique.\footnote{If it is not unique, we can define an $\epsilon$-regularized Lagrangian 
	and obtain an $\epsilon$-optimal solution.}
It is possible to identify a constant 
$\lambda_{\max}$ such that $\lambda_{\max}>\lamoptim$ \cite{MahdaviYJ2013}..
With this constant available, we set 
$\Lambda \eqdef [0,\lambda_{\max}]$.

Our first result is that $\optimal$ bounds the performance of any strategy in $\Strategy_\gamma$.
This result, as stated in  Theorem~\ref{lem:optimal}, is a generalization of the well-known result of  \cite{Algoet1994}  regarding the best possible performance for wealth alone (without constraints).

 \begin{thm}[Optimality of $\optimal$]
 \label{lem:optimal}
For any  investment strategy $\strategy \in  \Strategy_\gamma$ whose portfolios are $\bb_1,\bb_2,\ldots$, the following holds a.s.
\begin{gather*} 
\liminf_{T \rightarrow \infty} \frac{1}{T}\sum_{i=1}^T \omega (\bb_i,\obs_i)\geq \optimal.
\end{gather*}
 
 \end{thm}
From Theorem~\ref{lem:optimal} it follows that an investment strategy, $\Strategy \in \Strategy_\gamma$, is optimal if, for any bounded, stationary and ergodic process  $\{\obsF_i \}_{-\infty}^\infty$,
\begin{gather}
\label{eq:universal}
\lim_{T \rightarrow \infty} \frac{1}{T}\sum_{i=1}^T \omega (\bb_i ,\obs_i  ) = \optimal \phantom{a} a.s.
\end{gather} 
We find just such a strategy in Section~\ref{sec:Algorithm}.


%

\section{CVaR-Adjusted Nearest Neighbor Investment Strategy}
\label{sec:Algorithm}
\begin{algorithm}[tb!]
   \caption{CVaR-Adjusted Nearest Neighbor Investment Strategy (CANN)}
   \label{alg:main}
\begin{algorithmic}
\State \textbf{Input:} Countable set of experts $\{H_{k,h}\}$, $\alpha>0$
$(\bb_0,c_0)\in \choS$ $\lambda_{0}\in \Lambda$, initial probability $\{\beta_{k,h}\}$, 

\State \textbf{For $t=0$ to $\infty$}

\State \quad{}Play $\bb_{t},c_t,\lambda_{t}$. 

\State \quad{}Nature reveals market vector $\bX_{t}$

\State \quad{}Suffer loss $l(\bb_{t},c_t,\lambda_{t},\obsr_{t})$.
\State \quad{}Update the cumulative loss of the experts
\begin{gather*}
 l_{\cho,t}^{k,h}  \eqdef \sum_{i=0}^{t} l(\bb^{i}_{k,h},c^i_{k,h},\lambda_{i},\obsr_i)  \phantom{aaaa}  l_{\lambda,t}^{k,h}  \eqdef \sum_{i=0}^{t} l(\bb_{i},c_i,\lambda^{i}_{k,h},\obsr_i)
\end{gather*} 

\State \quad{}Update experts' weights  
\begin{gather*}
w_{t+1,\cho}^{(k,h)} \eqdef \beta_{k,h}\exp\left(-\frac{1}{\sqrt{t}}l_{\cho,t}^{k,h}\right) \\
p_{t+1,\cho}^{(k,h)} \eqdef \frac{w_{t+1,\cho}^{(k,h)}}{\sum_{h=1}^{\infty}\sum_{k=1}^{\infty}w_{t+1,\cho}^{(k,h)}}
\end{gather*}
\State \quad{}Update  experts' weights $w_{n+1}^{\lambda,(k,h)}$ 
\begin{gather*}
w_{t+1,\lambda}^{(k,h)}  \eqdef \beta_{k,h}\exp\left(\frac{1}{\sqrt{t}}l_{\lambda,t}^{k,h}\right) \\
p_{t+1,\lambda}^{(k,h)} \eqdef \frac{w_{t+1,\lambda}^{(k,h)}}{\sum_{h=1}^{\infty}\sum_{k=1}^{\infty}w_{t+1,\lambda}^{(k,h)}}
\end{gather*}
\State\quad{}Choose $\bb_{t+1},c_{t+1}$ and $\lambda_{t+1}$ as follows
\begin{gather*}
\bb_{t+1}  = \sum_{k,h}p_{t+1,\cho}^{(k,h)}\bb^{t+1}_{k,h} \phantom{aa} c_{t+1}  = \sum_{k,h}p_{t+1,\cho}^{(k,h)} c^{t+1}_{k,h} \phantom{aa} \lambda_{t+1} = \sum_{k,h}p_{t+1,\lambda}^{(k,h)} \lambda^{t+1}_{k,h}
\end{gather*} 
\State \textbf{End For}

\end{algorithmic}
\end{algorithm}

In this section we present an investment strategy in
$\Strategy \in \Strategy_\gamma$ that satisfies (\ref{eq:universal}). The strategy, which we call
\emph{CVaR-Adjusted Nearest Neighbor}, henceforth \algo, is summarized in the 
pseudo-code in Algorithm~\ref{alg:main}.
To define the strategy we require the following definition of 
the \emph{instantaneous Lagrangian}: 
\begin{equation}
\label{eq:l_loss}
l(\bb,c,\lambda,\obsr) \eqdef \omega(\bb,\obsr)+\lambda\left(c+\frac{1}{1-\alpha} \left( \omega (\bb,\obsr)-c\right)^+ -\gamma\right).
\end{equation}
The strategy  maintains a countable array of experts $\{H_{k,l}\}$,  where on each day $t$ an  expert $H_{k,l}$  outputs  a   triplet $(\bb^t_{k,l},c^t_{k,l} ,\lambda^t_{k,l})\in  \choS \times \Lambda$,
defined to be the minimax solution corresponding to an empirical distribution using nearest neighbor estimates
(see details below). 
We prove that, as $t$ grows, those empirical estimates  converge (weakly) to $\np_{\infty}$ and thus   converge to $\optimal$. 
Each day $t$, $\text{\algo}$ outputs a prediction $(\bb_t,c_t ,\lambda_t)\in  \choS \times \Lambda$. The sequence of predictions $(\bb_1,c_1),(\bb_2,c_2),\ldots$ output by $\text{\algo}$ is designed to minimize the average loss, 
$\frac{1}{T}\sum_{i=1}^T l(\bb,c,\lambda_i,\obsr_i)$.
Similarly, the sequence of predictions 
$\lambda_1,\lambda_2,\ldots$ is designed to   maximize the average loss, 
$\frac{1}{T}\sum_{i=1}^T l(\bb_i,c_i,\lambda,\obsr_i)$. 
Each of  $(\bb_i,c_i)$ and $\lambda_i$ is generated by aggregating the experts'
predictions $\cho^i_{k,l}$ and $\lambda^i_{k,l}$, $k,l=1,2,\ldots,$ respectively. In order to ensure that    $\text{\algo}$  
 will   perform as well as any other expert for both the $ \cho $ and   $\lambda$ predictions, we apply, twice simultaneously,  the Weak Aggregating Algorithm of \cite{Vovk2007}, and \cite{KalnishkanV2005}.   
It will also ensure that the average loss of the strategy will converge a.s. to $\optimal$.
 
We now turn to defining the  countable set of experts $\{H_{k,h}\}$: For  each $h = 1,2,\ldots$, we choose $p_h\in (0,1)$ such that  for the sequence $\{p_h\}_{h=1}^{\infty}$, $\lim_{h \rightarrow \infty} p_h =0$. 
Setting $\hat{h}=\lfloor np_h \rfloor$,
for expert $H_{k,h}$ we define, for a fixed  $k\times n$-dimensional vector, denoted $\stri$,  the following set, 
\begin{small}
\begin{gather*}
B^{\stri ,(1,n)}_{k,h}\eqdef \set{
\obsr_i \mid k+1\leq i \leq n, \obsF_{i-k}^{i-1} \text{ is among the } \hat{h}\phantom{a}   \text{nearest neighbors of}  \phantom{a} \stri \phantom{a} \text{among} \phantom{a} \obsF_{1}^{k},\ldots,\obsF_{n-k}^{n-1}},
\end{gather*}
\end{small}
where $\obsF_{j}^{j+k} \eqdef (\bX_j,\ldots ,\bX_{j+k})\in \nr^{k\times n}$.

Thus,  expert $H_{k,h}$ has a window of length $k$ and it looks for the $\hat{h}$ euclidean nearest-neighbors of $\stri$ in the past. We define also
\begin{gather*}
h_{k,h}^\cho (\obsF_{1}^{n-1},\stri) \eqdef    
\arg\min_{(\bb,c) \in \choS } \left(\max_{\lambda\in \Lambda} \frac{1}{|B^{\stri,(1,n)}_{k,h}|}\sum_{\obsr_i\in B^{\stri,(1,n)}_{k,h}}l_{k,l,n}(\bb,c,\lambda,\obsr_{i})\right) 
\\
h_{k,h}^\lambda (\obsF_{1}^{n-1},\stri) \eqdef  
\arg\max_{\lambda\in \Lambda} \left( \min_{(\bb,c) \in \choS} \frac{1}{|B^{\stri,(1,n)}_{k,h}|}\sum_{\obsr_i\in B^{\stri,(1,n)}_{k,h}}l_{k,l,n}(\bb,c,\lambda,\obsr_{i})\right)  
\end{gather*}
for 
\begin{gather*}
l_{k,h,n}(\bb,c,\lambda,\obsr_{i}) \eqdef 
 l(\bb,c,\lambda,\obsr_i)+\left(||(\bb,c)||^2-||\lambda||^2\right) \left(\frac{1}{n} +\frac{1}{h} +\frac{1}{k} \right),   
\end{gather*}
Using the above, we  define the predictions of $H_{k,h}$ to be:
\begin{gather}
H^\cho_{k,h} (\obsF_{1}^{n-1})=h^\cho_{k,h}(\obsF_{1}^{n-1}, \obsF_{n-k}^{n-1} ),\ n=1,2,3, \ldots \label{eq:expert1} \\
H^\lambda_{k,h} (\obsF_{1}^{n-1})=h^\lambda_{k,h}(\obsF_{1}^{n-1}, \obsF_{n-k}^{n-1} ), \label{eq:expert2}\ n=1,2,3, \ldots
\end{gather}
Note that $l_{k,h,n}(\bb,c,\lambda,\obsr)$ is  an approximation of $l(\bb,c,\lambda,\obsr)$, which guarantees that the minimax solution of every expert is unique. This technicality is used in the proof of Theorem~\ref{thm:Main}.

A $\gamma$-bounded investment strategy is called \emph{$\gamma$-universal} if its asymptotic average growth rate
is not worse than any $\gamma$-bounded strategy.
Theorem~\ref{thm:Main} below  states that the $\text{\algo}$ strategy, applied on  the experts defined above, is    $\gamma$-universal. We note that the theorem utilizes a standard assumption (see, e.g., \cite{BiauP2011,GyorfiUW2008}).
The proof of this theorem appears in the supplementary material.  
The main idea is to show first that the minimax (\ref{eq:minMax}) value of the Lagrangian (\ref{eq:Lagrangian}) is continuous with respect to the probability measure.  Then, we prove that the minimax measurable selection (which gives the 
optimal actions) is also continuous and every accumulation point of induced sequence of optimal actions 
is optimal.

\begin{thm}[$\gamma$-universality]
\label{thm:Main} 
 Assume that for any vector $\stri \in\nr^{n\times k}$ the random variable $||\obsF_1^k-\stri||$ has a continuous distribution. Then, for any $\gamma>0$ and for any bounded process 
 $\{\obsF_i\}_{-\infty}^{\infty}$, $\text{\algo}$ is $\gamma$-universal.
\end{thm}

\section{Empirical results}
\label{sec:empirical}

To apply the $\text{\algo}$ strategy, we implemented it with a finite set of experts, and 
in this section we present our empirical results on some standard datasets.
One objective of our experiments is to examine how well $\text{\algo}$  maintains the CVaR constrains. 
Another objective is to compare it to several well-known adversarial no-regret portfolio selection algorithms and to stochastically universal strategies. The benchmark algorithms we tested are:
\begin{itemize}
\item
		Best Constant Rebalancing Portfolio (BCRP)~\cite{Cover1991}: The BCRP is the optimal strategy in \emph{hindsight} 
		whenever market sequences are i.i.d.
\item Cover's Universal Portfolios (UP) \cite{Cover1991} , Exponentiated Gradient (EG)~\cite{HelmboldSSW1998}, Online Newton Steps (ONS)~\cite{AgarwalHKS2006}:  These  algorithms guarantee sub-linear regret w.r.t. the wealth achieved by BCRP.
		\item The nearest-neighbor based strategy (long-only and non-leveraged) of Gy{\"o}rfi et al. 		
		($\mathcal{B}_{NN}$)~\cite{GyorfiUW2008}: $\mathcal{B}_{NN}$, which is  a (stochastically) universal strategy whose asymptotic growth rate is optimal when the market follows a stationary and ergodic process.
		\item The nearest-neighbor based strategy (with short and leveraged): $\mathcal{B}^L_{NN}$ 
\end{itemize}
The experiments were conducted on two datasets that were used in many previous works (see, e.g., \cite{LiH2012,LiH2014,BorodinETG2004}). The first is the NYSE dataset, which consists of $23$ stocks between the years 1985-1995.
 The second is the MSCI dataset, which consists of $24$ stocks between the years 2006-2010. 
 Following \cite{GyorfiW2012,JohnsonB2015b}, for both datasets we used a daily interest rate of $r=0.000245$ and set $B=0.4$, which implies that $L_{B,r}=2.49$. While this interest rate is higher than the true rate in 2010, this choice only reduces the returns of our algorithm, which rarely  deposits cash and must pay a lot for short selling and loans.
 Similarly to the implementation of  $\mathcal{B}_{NN}$ \cite{GyorfiUW2008}, our implementation of $\text{\algo}$ took the following experts, $k=1,\ldots,5$ $h=1,\ldots,10$, for a total of $50$ experts,  and we set 
 $p_l = \frac{1}{20}+\frac{h-1}{18}$.  The initial expert prior  was set to be uniform and we chose the typical value of $\alpha=0.95$ for the calculation of CVaR. 
 The  hyper-parameters for the benchmark algorithms were according to \cite{OLPS}. 
\begin{table}[t]
	\caption{Wealth of $\text{\algo}$ and benchmark algorithms.}
	\label{tab:money-cvar}
	\vskip 0.15in
	\begin{center}
			\begin{sc}
				\begin{tabular}{l||llllll||l}
					\hline
					
					
					Dataset & BCRP & UP & EG  & ONS &  $\mathcal{B}_{NN}$ & $\mathcal{B}^L_{NN}$ & \algo$_{.05}$   \\
					\midrule

					NYSE     & $12.53$ & $5.05$  & $5.03$ & $5.83$ & $39.56$ & $1054$ & $58.8$ \\        
					 
					MSCI     & $1.51$ &$0.92$   & $0.93$ & $0.86$ & $13.47$ & $6.32$E$+05$ &$6.06$E$+03$  \\
					 
					\hline
				\end{tabular}
			\end{sc}
	\end{center}
	\vskip -0.1in
\end{table}

\begin{table}[t]
	\caption{$\cvar_{0.95}$ of $\text{\algo}$ with different values of $\gamma$.}
	\label{tab:cvar}
	\vskip 0.15in
	\begin{center}
		\begin{small}
		
			\begin{sc}
				\begin{tabular}{l||l|lllll}
					\hline
					
					
					Dataset & $\mathcal{B}^L_{NN}$ & \algo$_{.05}$ & \algo$_{.04}$  & \algo$_{.03}$ &  \algo$_{.02}$ & \algo$_{.01}$     \\
					\midrule

					NYSE     & $6.3\%$ & $3.2\%$  & $2.9\%$ & $2.46\%$ & $1.86\%$ & $1.24\%$  \\        
					 
					MSCI     & $7.76\%$ & $4.44\%$  & $3.81\%$ & $2.98\%$ & $2.27\%$ & $1.59\%$  \\
					 
					\hline
				\end{tabular}
			\end{sc}
		\end{small}
		
	\end{center}
	\vskip -0.1in
\end{table}

\begin{figure*}[t!]
    \centering
    \begin{subfigure}[t]{0.5\textwidth}
        \centering
        \includegraphics[height=1.8in]{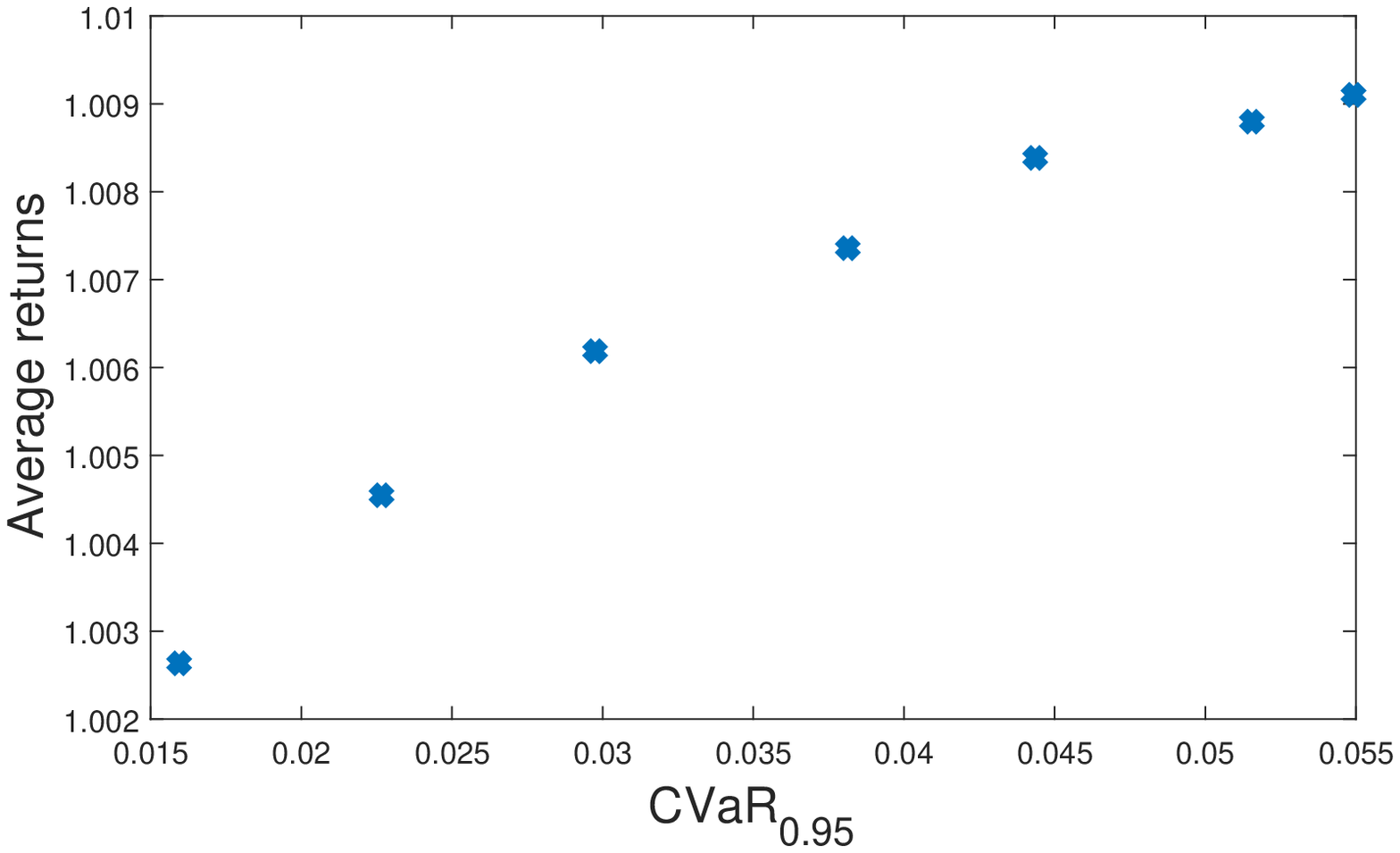}
        \caption{MSCI dataset}
    \end{subfigure}%
    ~
    \begin{subfigure}[t]{0.5\textwidth}
        \centering
        \includegraphics[height=1.8in]{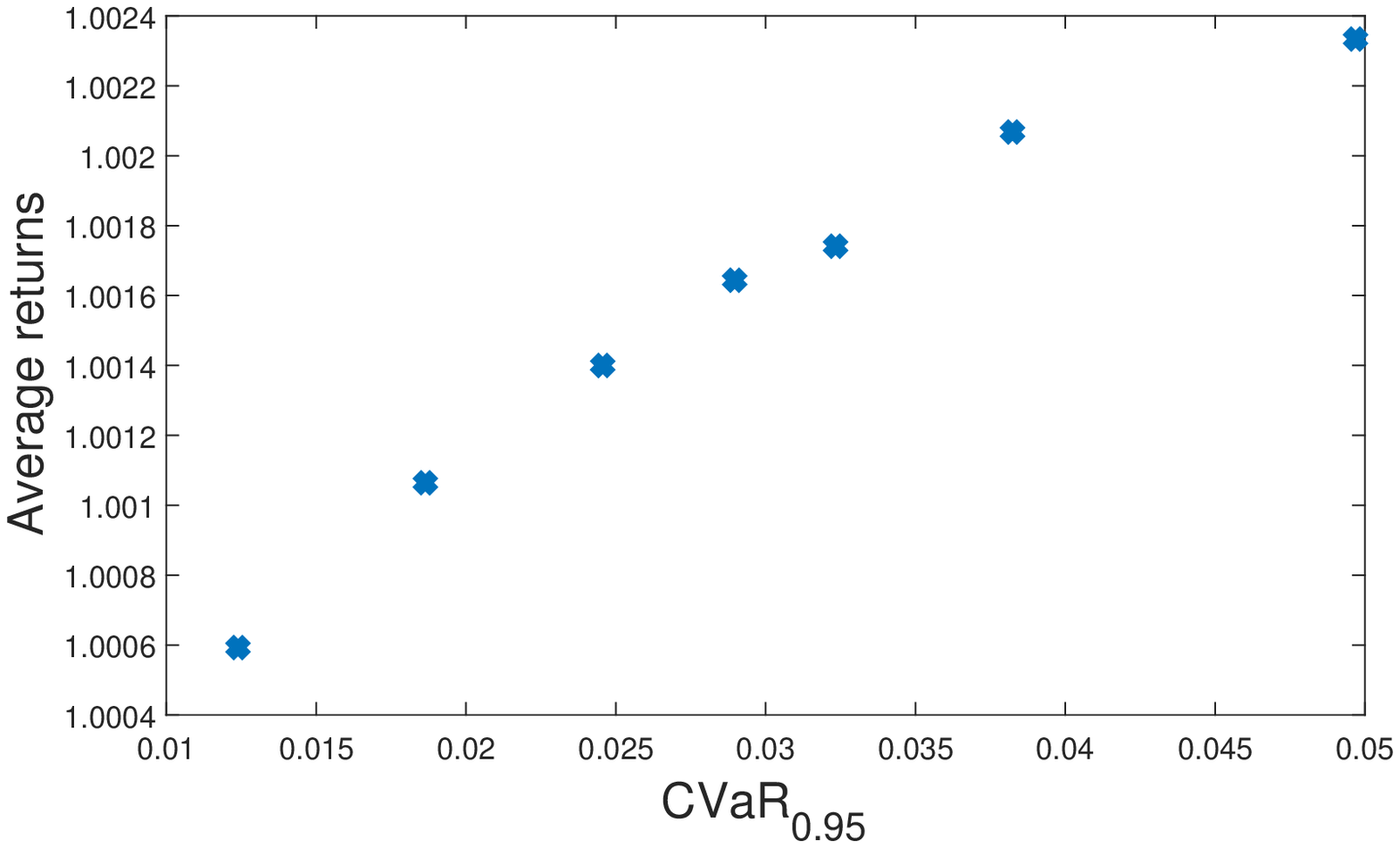}
        \caption{NYSE dataset}
    \end{subfigure}
    \caption{Mean-CVaR trade-off}
    \label{fig:mean-cvar}
\end{figure*}

\begin{figure*}[t!]
    \centering
    \begin{subfigure}[t]{0.5\textwidth}
        \centering
        \includegraphics[height=1.65in]{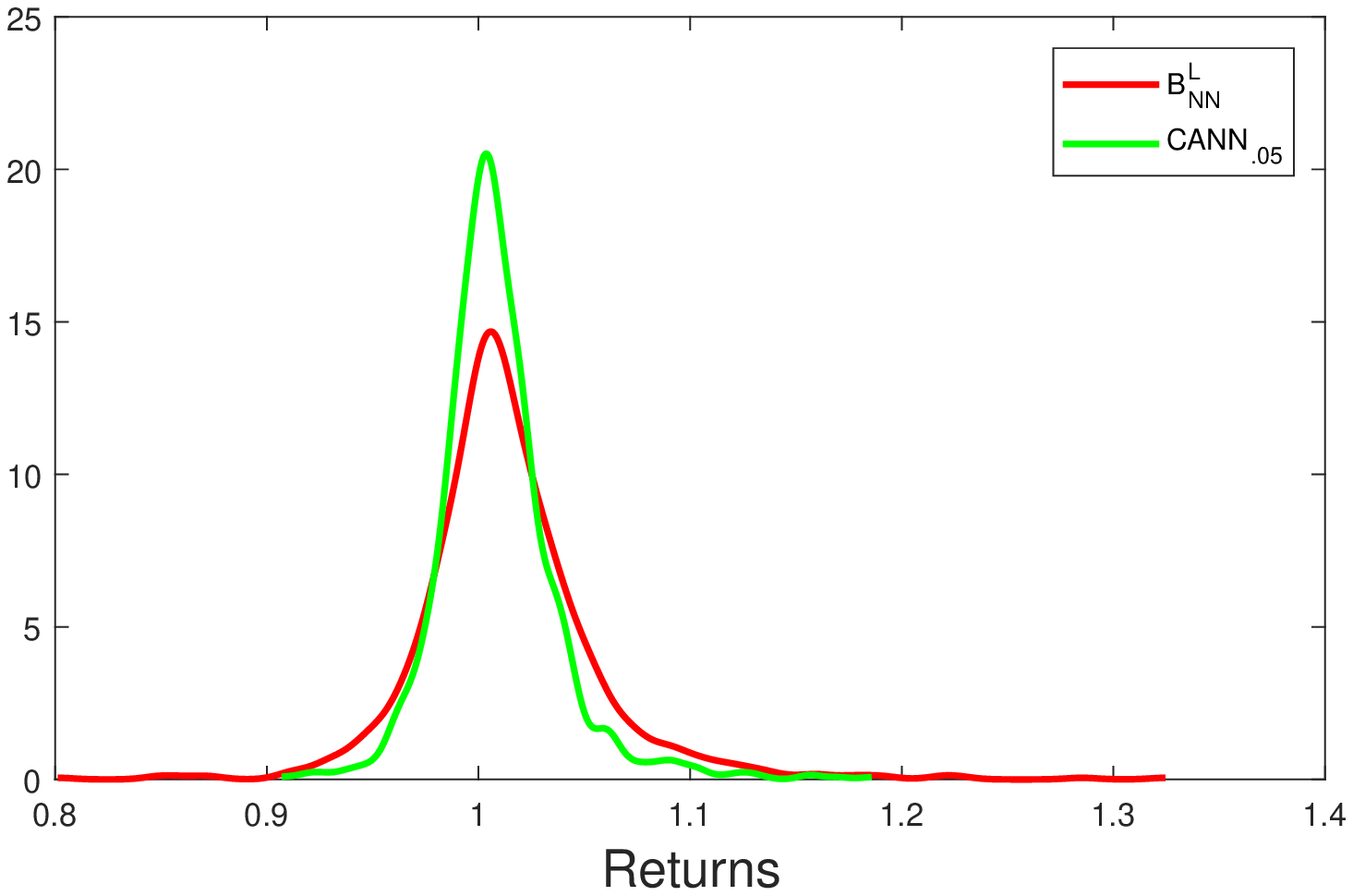}
        \caption{MSCI dataset}
    \end{subfigure}%
    ~
    \begin{subfigure}[t]{0.5\textwidth}
        \centering
        \includegraphics[height=1.65in]{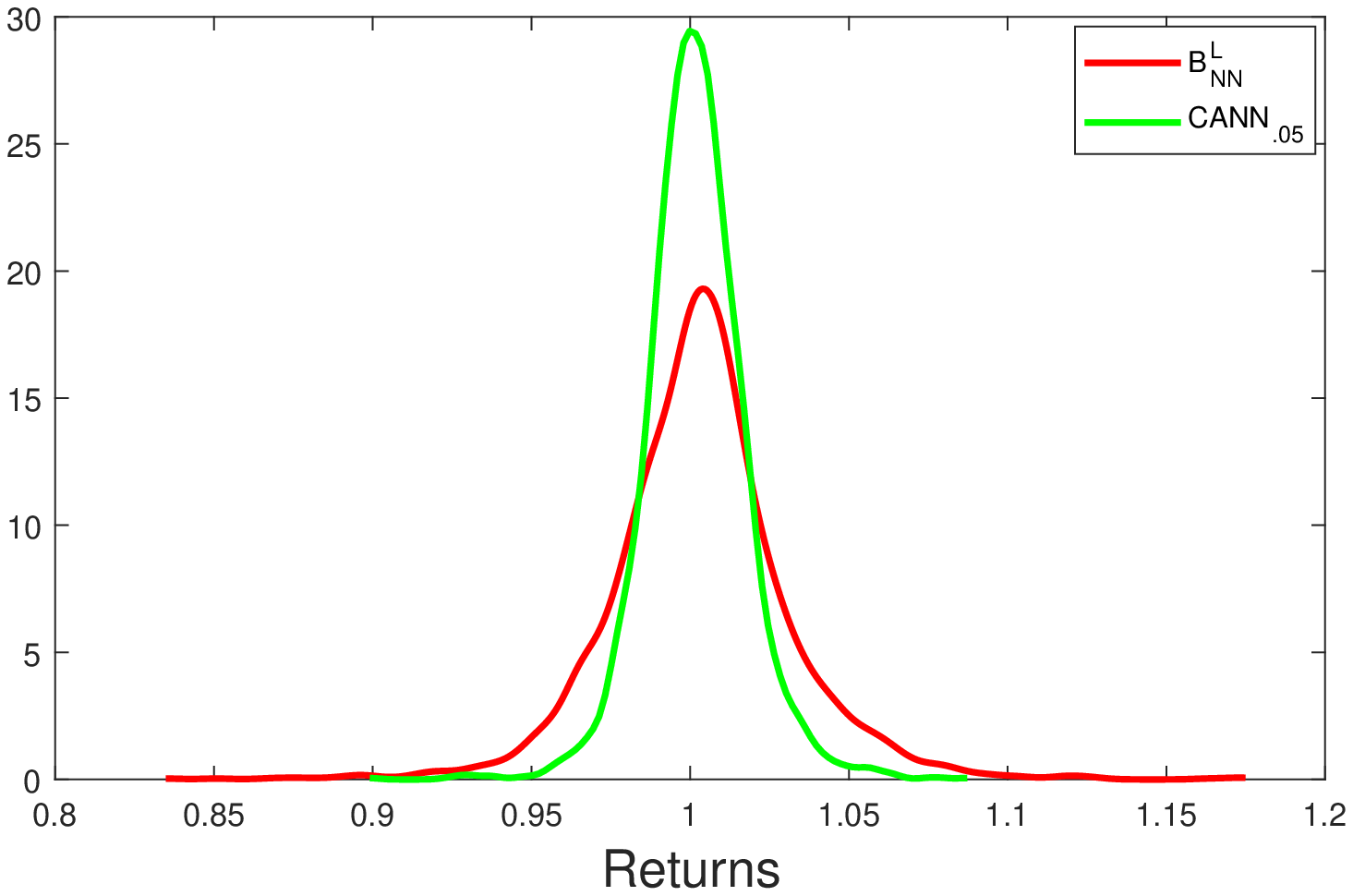}
        \caption{NYSE dataset}
    \end{subfigure}
    \caption{Empirical PDF}
    \label{fig:cdf}
    
\end{figure*}

Table~\ref{tab:money-cvar} presents the total wealth of all the algorithms, where $\text{\algo}$ was applied was $\gamma=0.05$. It is evident that the stochastically universal algorithms are superior to all the worst-case universal algorithms.   
In Figure~\ref{fig:cdf} we present the smoothed PDF of the returns of both $\mathcal{B}^L_{NN}$ and our algorithm.
The left tails of these PDFs show that our algorithm effectively decreases the losses.
 Another interesting aspect of our strategy is its   lower variance. We conducted another experiment where  we applied $\text{\algo}$ with different choices of $\gamma$ in the range $[0.01,0.07]$. The results are presented in Table~\ref{tab:cvar}, where the $\cvar_{0.95}$ is presented, and in Figure~\ref{fig:mean-cvar}, where the $y$-axis shows  the average return is presented and on the $x$-axis shows the CVaR$_{0.95}$. It can be seen that lower $\gamma$s result in  less risky strategies.
  Moreover, the concave shape suggests that by choosing an appropriate $\gamma$, one may achieve a better mean-CVaR trade-off.
%
%
%
%
%

%
%
%
%
%
%
%
%
%
%

\section{Concluding Remarks}

In this paper we  introduced the CVaR-adjusted nearest-neighbor portfolio selection strategy, which is the first CVaR-adjusted universal portfolio selection strategy  when the underlying market process is stationary and ergodic.
It should be noted that  it is possible to revise our method to work with 
other modern measures of risk such as the \emph{optimized certainty equivalent} \cite{BenT2007},
\emph{distortion risk measures} (mixture of CVaR)  \cite{FollmerS2002,Kusuoka2001}, and 
\emph{law-invariant coherent risk measures} \cite{Kusuoka2001}.

Early works in modern finance assumed that markets are stochastic and very simple (e.g., the returns are normally distributed) \cite{Markowitz1952,Malkiel1999}. This modeling assumption was later found to be too simplistic \cite{LoM2002}.
At the other extreme, Cover initiated the study of adversarial portfolio selection whereby stock prices are 
controlled by an adversary. Neither extreme led to overly effective strategies.
It appears that a more sophisticated stochastic modeling, as we pursue here, can lead to effective strategies; however, despite the empirical success of these methods, the bounds that can be obtained are asymptotic.
To overcome this  barrier, additional, and possibly strong, assumptions on the market process will be required. In the future, we wish  to pursue finite sample guarantees while not over-committing to dubious assumptions.    
\bibliographystyle{plain}
\bibliography{Bibliography}

\end{document}